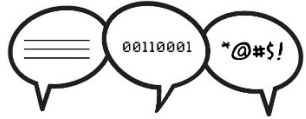

# Freedom of Speech and AI Output

*Eugene Volokh, Mark A. Lemley & Peter Henderson* [*]

## Introduction

Is the output of generative AI entitled to First Amendment protection? We're inclined to say yes. Even though current AI programs are of course not people and do not themselves have constitutional rights, their speech may potentially be protected because of the rights of the programs' creators. But beyond that, and likely more significantly, AI programs' speech should be protected because of the rights of their users—both the users' rights to listen and their rights to speak. In this short Article, we sketch the outlines of this analysis.

## I. AI Creators' Rights to Speak

AI programs aren't engaged in "self-expression"; as best we can tell, they have no self to express. They generate text or images in automated ways in response to prompts and based on their training. While people commonly anthropomorphize AI, speaking of it "memorizing" or "hallucinating" (and we've been guilty of that ourselves), the fact that AI generates text and images that we imbue with meaning doesn't mean that the AI is reasoning or even seeking to communicate with people.

---

[*] © 2023 Mark A. Lemley, Eugene Volokh, and Peter Henderson. Mark A. Lemley is the William H. Neukom Professor at Stanford Law School and of counsel at Lex Lumina PLLC. Eugene Volokh is the Gary T. Schwartz Professor of Law at UCLA School of Law. Peter Henderson is a J.D.-Ph.D. candidate in Computer Science and Law at Stanford University as well as incoming Assistant Professor in the Department of Computer Science and School of Public and International Affairs at Princeton University.

But someone creates AI programs, whether AI companies, universities, or people. AI creators' speech, like the speech of corporations or organizations generally, is protected by the First Amendment.[1] AI companies and their high-level employees may use autonomous agents and generative AI as a tool for guiding the creation of speech, consistently with their preferences and general ideologies.

AI programs' output is, indirectly, the AI company's attempt to produce the most reliable answers to user queries, just as a publisher may establish a newspaper to produce the most reliable reporting on current events.[2] The choices companies and their employees make about what sources to train on and what results to modify using human feedback directly or indirectly influence the output of their AIs. The analysis shouldn't change simply because this is done through writing algorithms, selecting training data, and then fine-tuning the models using human input[3] rather than hiring reporters or creating workplace procedures.

This is not necessarily an ideal outcome, because it may prove too much. Corporations take any number of actions that influence the content of their speech, from hiring and firing employees to making business decisions that leave more money available for speech; presumably those actions should not all be protected by the First Amendment. And there may be ways to draw distinctions between AI-generated speech and traditional corporate speech. For example, human agents of the corporation typically set policies and generate speech on behalf of the corporation. Humans are less directly involved in communicating via foundation models, which are generating text that the humans have not necessarily written or even seen.[4] And clearly some of the things ChatGPT or MidJourney generate aren't intended or even contemplated by the people who programmed or trained the system. Dan Burk has compared content generated by AI to interesting patterns in clouds or the sound we hear from a seashell: Even if we imbue them with meaning, that doesn't mean they were created to mean something.[5]

---

[1] *See* Citizens United v. FEC, 558 U.S. 310, 365 (2010) (holding that the government "may not suppress political speech on the basis of the speaker's corporate identity"); First Nat'l Bank v. Bellotti, 435 U.S. 765, 798 (1978) (holding that the First Amendment protects corporate speech). Corporate speech may be either traditionally protected speech that seeks to communicate information, which is entitled to full First Amendment protection, or commercial speech—speech that does no more than propose a commercial transaction. Commercial speech is also protected by the First Amendment, but receives less protection than traditionally protected speech, as we discuss below.

[2] *See* Eugene Volokh & Donald M. Falk, *First Amendment Protection for Search Engine Search Results*, 8 J. L. ECON. & POL. 883 (2012) (white paper commissioned by Google).

[3] *See, e.g.*, Long Ouyang (OpenAI) et al., *Training Language Models to Follow Instructions with Human Feedback*, https://perma.cc/6S3V-BZ48.

But that analysis may go too far in the other direction. AI models are clearly designed for the purpose of generating content that humans understand, and they are trained and refined to facilitate some forms of that content and inhibit others. That is even more true when we move from foundation models to models that are built on them. When small groups of people outside of companies—or even individuals—meticulously craft the speech generated by AI models to reflect their own personal views,[6] this may well be a conduit for those people's ideas. Yet drawing such boundaries based on the level of human involvement will inevitably fall into highly fact-specific inquiries and murky line-drawing.

The question whether the AI model creators' or deployers' First Amendment rights should extend to speech generated by the model—especially speech that the humans involved could not predict would be generated—may thus be hard. But it may prove unnecessary to answer, given the First Amendment rights of AI companies' users.

---

[4] *Cf.* Larry Lessig, *The First Amendment Does Not Protect Replicants*, in SOCIAL MEDIA, FREEDOM OF SPEECH, AND THE FUTURE OF OUR DEMOCRACY 275 (Lee Bollinger & Geoffrey Stone eds., 2022) (arguing that speech that "is crafted or originated algorithmically, with the substance of that algorithm not in any meaningful sense programmed by any individual in advance" shouldn't be treated as speech of those who created the algorithm); *see also* Karl M. Mannheim & Jeffery Atik, *White Paper: AI Outputs and the First Amendment* at 5 (working paper), https://papers.ssrn.com/sol3/papers.cfm?abstract_id=4524263 (arguing that AI output that doesn't merely adapt or report on the words of a human should be seen as unprotected by the First Amendment).

[5] *See* Dan L. Burk, *Asemic Defamation, or, The Death of the AI Speaker* (working paper).

[6] This is now common practice. *See, e.g.,* Yuntao Bai et al., *Constitutional AI: Harmlessness from AI Feedback* (2022) (manuscript), https://arxiv.org/abs/2212.08073 (researchers provided a "constitution" to the AI with instructions on how to respond to queries); Ouyang et al., *supra* note 3 (using human annotators to encourage certain outputs from a model aligning with the company's (OpenAI's) preferences for generated speech); Chunting Zhou et al., *Lima: Less Is More for Alignment* (2023) (manuscript), https://arxiv.org/abs/2305.11206 (handcrafting a detailed set of examples and tuning a model so it matches the style and preferred response strategies of the model creator team).

## II.	Users' Rights to Listen

We think the strongest argument for First Amendment protection stresses not the AI or its corporate owner as speaker but rather the interests of listeners in receiving meaningful communication. The First Amendment protects "speech" and not just speakers; and while the Fourteenth Amendment protects the liberty of "person[s]," that includes the liberty to hear and not just to speak.[7]

To the extent that the First Amendment aims to protect democratic self-government, the search for truth, and the marketplace of ideas, that must extend to the rights of those who would consider the speech in making democratic decisions, in trying to identify the truth, and in weighing the value of rival ideas, and not just to the rights of those who create and distribute the speech. And to the extent that the First Amendment aims to protect the interests of thinkers or democratic citizens,[8] which wouldn't include AI programs and which might not include AI companies, listeners have their own rights as thinkers and democratic citizens, including the right to receive speech.[9] As users (for better or worse) increasingly turn to AI systems to easily access and aggregate information,[10] these core First Amendment concerns justify protecting such sources of information.

---

[7] *See also* Cass R. Sunstein, *Artificial Intelligence and the First Amendment*, https://papers.ssrn.com/sol3/papers.cfm?abstract_id=4431251 (working paper) (taking a similar view).

[8] *See, e.g.*, C. Edwin Baker, Human Liberty and Freedom of Speech 196, 204 (1989); Robert Post, *Participatory Democracy and Free Speech*, 97 Va. L. Rev. 477 (2011); Seana Valentine Shiffrin, *A Thinker-Based Approach to Freedom of Speech*, 27 Const. Comment. 283, 297 (2011); James Weinstein, *Participatory Democracy as the Central Value of American Free Speech Doctrine*, 97 Va. L. Rev. 491 (2011).

[9] *See* Jane Bambauer, *Negligent AI Speech: Some Thoughts About Duty*, 3 J. Free Speech L. 343, 347–48 (2023). While some arguments for free speech may stress the value of hearing arguments "from persons who actually believe them; who defend them in interest, and do their very utmost for them," John Stuart Mill, On Liberty 72 (1863), that isn't a basis for stripping protection from, say, speech by the devil's advocate or the impartial recounter—human or algorithmic—of others' assertions.

[10] For instance, since ChatGPT's launch, researchers have found that StackOverflow, a popular forum for getting answers to coding-related questions, has lost a significant amount of its traffic. Maria del Rio Chanona, Nadzeya Laurentsyeva & Johanned Wachs, *Are Large Language Models a Threat to Digital Public Goods? Evidence from Activity on Stack Overflow* (2023) (manuscript), https://arxiv.org/pdf/2305.11206.

Indeed, the Court has long recognized First Amendment rights "to hear" and "to receive information and ideas."[11] Regardless of whether any speaker interests are involved in an AI program's output, readers can gain at least as much from what the program communicates as they do from commercial advertising, corporate speech, and speech by foreign propagandists—three kinds of speech that have been held to be protected in large part because of listener interests.[12] Courts also

---

[11] *See* Kleindienst v. Mandel, 408 U.S. 753, 762–63 (1972) ("In a variety of contexts this Court has referred to a First Amendment right to 'receive information and ideas' . . . ."); Stanley v. Georgia, 394 U.S. 557, 564 (1969) ("It is now well established that the Constitution protects the right to receive information and ideas."); Thomas v. Collins, 323 U.S. 516, 534 (1945) ("That there was restriction upon Thomas' right to speak and the rights of the workers to hear what he had to say, there can be no doubt."); Red Lion Broadcasting Co. v. FCC, 395 U.S. 367, 386–90 (1969) ("It is the right of the public to receive suitable access to social, political, esthetic, moral, and other ideas and experiences which is crucial here.").

[12] Va. State Bd. of Pharmacy v. Va. Citizens Consumer Council, Inc., 425 U.S. 748, 756 (1976) (concluding that commercial speech is protected because "protection afforded is to the communication, to its source and to its recipients both"); *id.* at 757 ("in *Procunier v. Martinez*, 416 U. S. 396, 408–409 (1974), where censorship of prison inmates' mail was under examination, we thought it unnecessary to assess the First Amendment rights of the inmates themselves, for it was reasoned that such censorship equally infringed the rights of noninmates to whom the correspondence was addressed"); First Nat'l Bank of Boston v. Bellotti, 435 U.S. 765, 775–76, 783 (1978) (concluding that corporate speech is protected "based not only on the role of the First Amendment in fostering individual self-expression but also on its role in affording the public access to discussion, debate, and the dissemination of information and ideas"); Lamont v. Postmaster Gen., 381 U.S. 301, 305, 307 (1965) (relying on "the addressee's First Amendment rights" rather than the sender's, where the sender was a foreign government); *see also id.* at 307–08 (Brennan, J., concurring) (stressing that it's not clear whether the First Amendment protects "political propaganda prepared and printed abroad by or on behalf of a foreign government," but concluding that the law was unconstitutional because it violates the recipients' "right to receive" information, regardless of the senders' rights to speak); Toni M. Massaro, Helen Norton & Margot E. Kaminski, *SIRI-Ously 2.0: What Artificial Intelligence Reveals About the First Amendment*, 101 Minn. L. Rev. 2481, 2487–91 (2017); Toni M. Massaro & Helen Norton, *Siri-Ously? Free Speech Rights and Artificial Intelligence*, 110 Nw. U. L. Rev. 1169, 1176–78, 1183–85 (2016). To be sure, the rights of corporations and commercial speakers are also justified in part by speaker interests; our point here is that the Court has stressed listener interests as an important (and potentially independent) basis for protection as well.

Larry Lessig argues that *Bluman v. FEC*, 800 F. Supp. 2d 281 (D.D.C. 2011), *aff'd*, 565 U.S. 1104 (2012), which upheld a prohibition on campaign contributions and expenditures by certain noncitizens, "mean[s] that the fact that there are listeners does not elevate all political speech into protected political speech." *Lessig*, *supra* note 4, at 281. But *Bluman* stems from the special rules that the Court periodically hold that dead people have no constitutional rights,[13] but there is certainly a First-Amendment-protected interest in reading the writings of, say, Aristotle or Galileo or Machiavelli.[14]

Consider, for instance, a state law restricting AI output that is critical of the government, or that discusses abortion or gender identity or climate change (even if the restriction is framed in an ostensibly viewpoint-neutral way). Such a law would undermine users' ability to hear arguments that they might find persuasive and relevant to their political and moral decisionmaking. The First Amendment should protect readers from such government restrictions on speech composed by AI programs, just as it protects readers from government restrictions that block the readers from seeing speech composed by foreign governments or corporations or dead authors.

Commercial advertising is less protected from speech restrictions than other speech, especially when it is false or misleading; but this stems from other features

---

has developed for foreigners' participation in specific governmental processes (including election campaigns). 800 F. Supp. 2d at 288–89. It doesn't undermine the holding of *Lamont*, under which the existence of American listeners does elevate foreigners' speech—even speech sent into the U.S. from foreign countries by foreign governments—into speech protected against American-government-imposed restraints.

[13] The statements are general, though the applications mostly involve constitutional rights other than the First Amendment. Whitehurst v. Wright, 592 F.2d 834, 840 (5th Cir. 1979) ("After death, one is no longer a person within our constitutional and statutory framework, and has no rights of which he may be deprived."); *see also* Silkwood v. Kerr-McGee Corp., 637 F.2d 743, 749 (10th Cir. 1980) ("[T]he civil rights of a person cannot be violated once that person has died."); Hillspring Health Care Ctr., LLC v. Dungey, No. 1:17-cv-35, 2018 WL 287954, at *10 (S.D. Ohio Jan. 4, 2018) ("a deceased person has no civil rights that may be violated"); State v. Powell, 497 So. 2d 1188, 1190 (Fla. 1986) ("[A] person's constitutional rights terminate at death."). For criticism of that rule, see Fred O. Smith, Jr., *The Constitution After Death*, 120 COLUM. L. REV. 1471 (2020).

[14] *See* Jack Balkin, *Artificial Intelligence and the First Amendment* (working paper 2023); Eugene Volokh, *Speech Restrictions That Don't Much Affect the Autonomy of Speakers*, 27 CONST. COMM. 347 (2011). We thus disagree with Mannheim & Atik, *supra* note 4, at 6, that the right of listeners to hear is limited to "human generated communications." Listeners' rights to hear are founded on the value of the communication to the listeners. That value can be present even if the communication was created by a speaker who lacks First Amendment rights—for instance, a foreign government, or someone who is long dead. And that value can be present even if the communication was created by an AI program, whether or not one thinks the AI program's creators are expressing themselves through the program.

of commercial advertising, not from its being justified by listener interests.[15] It is possible that certain specific types of AI program output—such as the output of AI systems that are highly integrated into personalized advertising delivery systems—are indeed a form of commercial advertising.[16] But there is little reason to apply commercial speech doctrine when AI programs produce material that has no connection to advertising. And speech outside commercial advertising does not lose protection because it is distributed for profit.[17]

### III.    USERS' RIGHTS TO SPEAK

People also often use AI programs to create their own speech: to create text that they can edit and then publish; to fill in gaps in already-drafted work; and, more indirectly, to get a general research summary of a subject that facilitates more research and then future speech. Some such uses may of course be dishonest, for instance if a student turns in an AI-written project in a course where such technological assistance is forbidden. But all technologies can be used dishonestly, and much AI-assisted writing is perfectly legitimate.

Courts have recognized that the First Amendment protects the use of technology to gather information and create materials that will be part of the users' speech. This is particularly clear in the recent circuit court cases recognizing the right to videorecord and audiorecord in public places (at least when people are recording government officials, such as police officers), because such recording enables further speech by the recorder.[18] The recorder isn't speaking; the recording itself doesn't directly facilitate listening; but the recording is constitutionally valuable because it can be communicated to others.[19]

---

[15] The Court's explanation for the lower level of protection for commercial advertising—as articulated in *Virginia Pharmacy*, the case that first squarely held that such advertising is generally protected—was that (1) "[t]he truth of commercial speech, for example, may be more easily verifiable by its disseminator than, let us say, news reporting or political commentary" and that (2) "[s]ince advertising is the sine qua non of commercial profits, there is little likelihood of its being chilled by proper regulation." 425 U.S. at 771 n.24. These elements—verifiability and durability—are likely not especially present for most AI output.

[16] *See* Massaro, Norton & Kaminski, *supra* note 12, at 2519 (discussing the argument that AI output should generally be treated more like commercial advertising).

[17] *See, e.g.*, Bd. of Trustees of State Univ. of New York v. Fox, 492 U.S. 469, 482 (1989) (noting that various "examples . . . of speech for a profit" "do not consist of speech that proposes a commercial transaction, which is what defines commercial speech," and noting that "[s]ome of our most valued forms of fully protected speech are uttered for a profit").

The same may be said about AI programs, even apart from users' rights as readers and (perhaps) AI companies' rights as speakers. They are tools for creating user speech, and therefore as protected by the First Amendment as are cameras and voice and video recorders.

All this is just a special case of the proposition that the First Amendment protects technologies that make it easier to speak. The "press" itself refers to one such technology, the printing press, which was of course both immensely valuable and immensely disruptive.[20] Since then, the Court has recognized such protection for film, cable television, the Internet, social media, and more.[21] The same should apply to generative AI. Just as the Internet and social media have become "the most important places . . . for the exchange of views," and are thus fully protected by the

---

[18] *See, e.g.*, Glik v. Cunniffe, 655 F.3d 78, 82 (1st Cir. 2011); Turner v. Driver, 848 F.3d 678, 689 (5th Cir. 2017); ACLU of Illinois v. Alvarez, 679 F.3d 583, 595–608 (7th Cir. 2012); Fordyce v. City of Seattle, 55 F.3d 436, 439 (9th Cir. 1995); Smith v. City of Cumming, 212 F.3d 1332, 1333 (11th Cir. 2000).

[19] *See, e.g.*, *ACLU of Ill.*, 679 F.3d at 607 ("audio and audiovisual recording are uniquely reliable and powerful methods of preserving and disseminating news and information about events that occur in public"). This also responds, we think, to the argument that material that is created by an AI program is unprotected by the First Amendment because it isn't created by a human. Mannheim & Atik, *supra* note 4, at 7. A video or audio recorder isn't a human, but its output is valuable to its user, because it facilitates the user's own speech—for instance, by letting the user distribute a videorecording of police abuse. Likewise with an AI program. (Of course, the user of the video or audio recorder generally decides what and when to record, but likewise a user decides what prompt to give an AI program.)

[20] *See* Eugene Volokh, *Freedom for the Press as an Industry, or for the Press as a Technology?*, 160 U. PA. L. REV. 459 (2012); FRANCIS LUDLOW HOLT, THE LAW OF LIBEL 38 (photo. reprint 1978) (1812) (defining "[t]he liberty of the press" as "the personal liberty of the writer to express his thoughts in the more [im]proved way invented by human ingenuity in the form of the press"; the original reads "the more approved," but the author must have meant "the more improved," and the 1816 revision makes that correction, FRANCIS LUDLOW HOLT, THE LAW OF LIBEL 51 (London, J. Butterworth et al. 2d ed. 1816)).

[21] Television and radio broadcasting remains a narrow technological exception to full First Amendment protection, but the Court has resisted extending that exception. *See, e.g.*, Reno v. ACLU, 521 U.S. 844, 868–70 (1997).

First Amendment, so AI programs are likely to be among the most important tools for people to be able to speak (as well as to listen).

## Conclusion

There are, of course, many nuances and potential challenges here. It's possible, for instance, that AI speech is generally fully protected but potentially subject to slightly more restrictions in certain situations. Perhaps some disclosure requirements may be constitutional because of the special nature of AI speech, for instance, as they appear to be constitutional for corporate speech[22] and for commercial speech.[23] Indeed, to the extent the protection for AI speech stems from the interests of listeners, providing more information for listeners may fit well with that protection.

Nonetheless, as a general matter, speech created by AI programs, like speech created by foreign governments, speech created by corporations, and speech created by the long-dead, is likely constitutionally protected—because of the First Amendment rights of those who would receive the speech, whether or not AI companies' own free speech rights are implicated as well.

---

[22] *See* Citizens United v. FEC, 558 U.S. 310, 369–71 (2010) (8–1 on this point).

[23] Zauderer v. Office of Disciplinary Counsel, 471 U.S. 626, 651 (1985).